# DFT calculations of point defects on UN(001) surface

D. Bocharov[1, 2, 3], D. Gryaznov[1], Yu.F. Zhukovskii[1], and E.A. Kotomin[1]

[1]*Institute for Solid State Physics, Kengaraga 8, LV-1063 Riga, Latvia*
[2]*Faculty of Physics and Mathematics, University of Latvia, Zellu 8, LV-1002 Riga, Latvia*
[3]*Faculty of Computing, University of Latvia, Raina blvd 19, LV-1586 Riga, Latvia*

**Abstract.** The density functional theory is used in a study of point defects on both UN (001) surface and sub-surface layers. We compare the results for slabs of different thicknesses (both perfect and containing nitrogen or uranium vacancies) with a full geometry, electronic and spin density optimization. The electronic charge density re-distribution, density of states, magnetic moments of U atoms and local atomic displacements around defects are carefully analyzed. It is predicted that the vacancies are formed easier on the surface whereas the property of sub-surface layer does not differ significantly from the central one in the slab.

## 1. Introduction

Uranium mononitride (UN) is considered nowadays by the Generation IV International Forum of nuclear reactors [1] as one of the promising nuclear fuels alternative to $UO_2$. However, it reveals unwanted oxidation in air [2] which could affect the fuel fabrication process and its performance. Atomistic understanding of the oxidation process could help to solve this problem.

Previous first-principles simulations on UN used mostly the density functional theory (DFT) and were focused mainly on bulk properties (for example, [3-9]). To check reliability of these results, we performed recently several calculations on bulk and (001) surface of UN using the two different DFT approaches [10]: linear combination of atomic orbitals (LCAO) applied for construction of basis sets and plane waves (PW) combined with the pseudopotentials representing the core electrons, as implemented in both *CRYSTAL* [11] and *VASP* [12] computer codes. Our basic findings for the bulk and the (001) surface of UN calculated using the *VASP* code were confirmed by *CRYSTAL* calculations [10]. The results of both series of calculations on the lattice constant, bulk modulus, cohesive energy, charge distribution, band structure and density of states (DOS) for UN single crystal were analyzed.

Recently [13, 14], we performed first principles simulations on the atomic and molecular oxygen interaction with the perfect UN(001) surface. It was demonstrated that the $O_2$ molecules could spontaneously dissociate [14] at the defect-free surface and releasing O adatoms reveal strong chemical interaction with surface ions [13]. It is worth mentioning that all our UN surface calculations [10, 13, 14] were performed for the fixed magnetic moments of U atoms.

Corresponding author: Fax.:+371 67132778
E-mail address: bocharov@latnet.lv (D. Bocharov).

To understand the oxidation mechanism in more detail, one has to take into account *surface defects* and their interaction with oxygen. So far, only point defects in the UN bulk were calculated [15, 16]. In this paper, we study basic properties of *surface* vacancies. In Section 2, a slab model and parameters used in our present spin-polarized PW DFT calculations are described. In Section 3, we discuss main results obtained for the N and U vacancies on the surface. A short summary is presented in Section 4.

**2. Slab model and computational details**

UN possesses a rock-salt *fcc* structure over a wide temperature range. We model the (001) surface using the symmetrical slabs containing odd number (5, 7, 9 or 11) of atomic layers separated by the vacuum gap of 38.9 Å which corresponds to 16 inter-layers (Fig. 1). Atomic layers consist of regularly alternating N and U atoms. Our test calculations show that such an inter-slab distance is large enough to exclude spurious interactions between the slabs repeated in the *z*-direction.

To simulate single point defects (either N or U vacancies), we applied a supercell approach using unit cells with 2×2 and 3×3 extensions of surface translation vectors. These supercells contain four and nine pairs of atoms in each layer while periodically distributed surface vacancies for such unit cells correspond to defect concentrations of 0.25 and 0.11 monolayers (ML), respectively. We calculated not only the outer surface defects, but also the sub-surface defects as well as those positioned at the central layer of the slab. To reduce computational efforts, we considered the two-side arrangement of the point defects which is symmetrical with respect to the central (mirror) plane (the atomistic model of surface N vacancies with the 2×2 periodicity is shown in Fig. 2).

For calculations, we used the PW DFT computer code *VASP 4.6* [12,17]. To represent the core electrons (78 electrons for U and 2 electrons for N), the relativistic pseudopotentials combined with the PAW method [18] were used. The Perdew-Wang-91 non-local exchange-correlation (GGA) functional [19] was chosen for calculations. The cut-off energy was fixed at 520 eV. The Monkhorst-Pack *k*-point mesh [20] of 8×8×1 for integration over the Brillouin zone (BZ) was used whereas the electron occupancies were determined following Methfessel and Paxton [21] as implemented in the *VASP* code. The smearing parameter of 0.2 eV was found to be optimal for reasonable convergences suggesting the electronic entropy contribution of the order of 10 meV. The total energy of slabs of different thicknesses was optimized with respect to the atomic positions only, with the lattice parameter fixed at its equilibrium value of 4.87 Å for UN bulk. This value is slightly underestimated as compared to the experimental bulk value of 4.89 Å [22]. The ferromagnetic state was chosen for all our slab calculations [3] performed for the self-consistent (relaxed) atomic magnetic moments with no spin-orbit interactions included. Consequently, we calculated both the effective atomic charges and average magnetic moments *per* atom using the topological Bader analysis [23, 24].

**3. Main results**

*3.1. Perfect UN(001) surface*

First, the calculations of the effective atomic charges $q^{eff}$, atomic displacements $\Delta r$, average magnetic moments $\mu_{av}$ of U atoms, and surface energies $E_{surf}$ for defect-free slabs of different thicknesses (Tables 1 and 2) were performed, in order to check how these properties depend on atomic spin relaxation (in our previous calculations the total magnetic moment of a slab was fixed [10, 13, 14]). The spin relaxation leads to considerable change of the $E_{surf}$ depending on the number of layers in a slab (Table 1). The largest $\mu_{av}$ value was obtained for the U atoms in the 5-layer slab, *i.e.*, $\mu_{av}$ slightly decreases with the thickness suggesting difference of 0.3 $\mu_B$ between the 5- and 11-layer slabs. The lattice relaxation energies in spin-optimized calculations turn out to be quite small, *i.e.,* ~0.03 eV.

It is also interesting to analyze $q^{eff}$ values for atoms across the slab as a function of the number of layers in a slab (Table 2). First, $q^{eff}$ shows considerable covalent bonding both on the surface (*e.g.*, sub-surface) and on the central plane since the values are quite far from the formal ionic charges ±3*e*. Second, the values in Table 2 demonstrate that the surface is slightly positively charged, due to a difference in the N and U charges. Third, the atomic charges are insensitive to both the spin relaxation and the number of layers.

The atomic displacements *Δz* from perfect lattice sites differ significantly for U atoms positioned at the surface and sub-surface layers (Table 3) being somewhat larger for the 5-layer slab while displacements of nitrogen atoms for all the slabs remain almost unchanged. Note that N atoms at (001) surface are displaced up whereas U atoms are shifted inwards the slab center which results in the surface rumpling up to 1.2% of the lattice constant.

### 3.2. Vacancies on the (001) surface

In the present study, we considered the two reference states in calculations of the defect formation energies, both widely used in the literature. The point defect formation energy was calculated either as

$$E_{form}^{N(U)vac} = \frac{1}{2}\left(E_{def}^{UN} + 2E_{ref\_I(II)}^{N(U)} - E^{UN}\right), \quad (1a)$$

for surface and sub-surface vacancies, or

$$E_{form}^{N(U)vac} = E_{def}^{UN} + E_{ref\_I(II)}^{N(U)} - E^{UN}, \quad (1b)$$

for a vacancy in the central layer of the slab. Here $E_{def}^{UN}$ is the total energy of fully relaxed slab containing N (or U) vacancies, $E^{UN}$ the same for a defect-free slab, while $E_{ref\_I(II)}^{N(U)}$ is reference energy for N (or U) atom. In our study, we used the two different reference states for both N and U atoms.

The first reference corresponds to N (U) isolated atom in triplet (quartet) spin states determined by $2p^3$ ($5f^{\,3}6d^1$) valence electron configurations (hereafter reference I as in Table 4) calculated in a large tetragonal box (28.28×28.28×22 Å$^3$), *i.e.*:

$$E_{ref\_I}^{N(U)} = E_{atom}^{N(U)} \quad (2)$$

The second reference state (hereafter reference II as in Table 4) represents the chemical potential of N (U) atom which is in general a function of temperature and nitrogen partial pressure. By neglecting these effects, the N chemical potential can be treated as the energy of atom in the molecule $N_2$. Consequently, the chemical potential of U atom is given by the one-half total energy (per unit cell) of U single crystal in its low-temperature α-phase having the orthorhombic structure [25]. Thus, the corresponding second reference energies can be estimated as:

$$E_{ref\_II}^{N} = m_{N_2} = \frac{1}{2} E_{tot}[N_2],$$  (3a)

$$E_{ref\_II}^{U} = m_{\alpha\text{-}U} = \frac{1}{2} E_{tot}[\alpha\text{-}U],$$  (3b)

where $E_{tot}[N_2]$ is the total energy of nitrogen molecule while $E_{tot}[\alpha\text{-}U]$ the total energy of U bulk unit cell containing two atoms. The chemical potentials of N and U, as calculated according to Eq. 3, represent extreme cases of N (U) - rich conditions [26], *i.e.*, their minimum values were not considered in the present study. The formation energy of N (U) vacancy with respect to the $N_2$ molecule (or α-U single crystal) and the energy of N (U) isolated atom are closely related: the former being larger than the latter by half the binding energy of the $N_2$ molecule or half the cohesive energy of α-U single crystal.

The lattice parameters of α-U were optimized: $a$ = 2.80 Å, $b$ = 5.88 Å, $c$ = 4.91 Å which are slightly underestimated as compared to values obtained experimentally [25] and calculated elsewhere [27, 28], except for the parameter $b$ which is in a good agreement with experimental value of 5.87 Å [25] (while $a$ = 2.86 Å, $c$ = 4.96 Å [25]). Also, the ratios $c/a$, $b/a$ as well as the parameter $c$ are well verified by another plane-wave DFT study [29]. Analogously to an isolated nitrogen atom, the $N_2$ molecule was calculated in the box but of a smaller size (8×8×8 Å$^3$). The molecule $N_2$ is characterized by the bond length of 1.12 Å and the binding energy of 10.63 eV being qualitatively well comparable with the experimental values of 1.10 Å and 9.80 eV [30], respectively. Note that the pre-factor of ½ in Eq. (1a) arises due to a mirror arrangement of two N (U) vacancies on the surface and sub-surface layers (Fig. 2).

The formation energies of N and U vacancies ($E_{form}^{N(U)\,vac}$) calculated using Eqs. (1-3) (with the two reference states as functions of the slab thickness) are collected in Table 4. These are smallest for the surface layer and considerably increase by ~0.6 eV for the N vacancy and by ~1.7 eV for the U vacancy in the sub-surface and central layers, independently of the reference state. This indicates the trend for vacancy segregation at the interfaces (surface or grain boundaries). A weak dependence of $E_{form}^{N(U)\,vac}$ on the slab thickness is also observed. The value of $E_{form}^{N(U)\,vac}$ is saturated with the slab thicknesses of seven atomic layers and more. Moreover, the difference between values of $E_{form}^{N(U)\,vac}$ for the 5 and 7 layer slabs is less for the surface vacancies than for those in the central layer. This difference is the largest for the U vacancy in the central plane (~0.16 eV).

The reference state II leads to smaller $E_{form}^{N(U)\,vac}$ (as compared to those found with the reference state I) and demonstrates a significant difference for two types of vacancies. According to reference II, the U vacancy could be substantially easier formed at T = 0 K than the N vacancy. Notice that the chemical potentials of O and U atoms used in similar defect studies on $UO_2$ bulk did not reveal the energetic preference for U vacancy [28, 31]. The defect-defect interaction is not responsible for this effect as $E_{form}^{N(U)\,vac}$ decreased by 0.1 eV only with the larger supercell size (3×3 in Table 4). On the other hand, due to the temperature dependence of the chemical potential of a free $N_2$ molecule [32], we predict reduction of the formation energy of the N vacancy by ~0.8 eV as the temperature increases from RT to 1000 $^o$C. Unlike the reference state II, the reference I results in similar formation energies for both types of the vacancies. In the central slab layer, values of $E_{form}^{N(U)\,vac}$ were found to be similar to those in the bulk (Table 4).

The local atomic displacements around the vacancies are largest for the nearest neighbors of vacancies. The analysis of atomic displacements allows us to suggest that the U vacancy disturbs the structure of the surface stronger than the N vacancy. If the N vacancy lies in the surface layer, displacements of the nearest U atoms in $z$-direction achieve 0.02-0.05 Å towards the central plane of slab. The displacements of N atom nearest to surface N vacancy achieve 0.05 Å towards the central plane ($z$-direction) and 0.01 Å in $xy$ (surface) plane. Maximum displacements of neighbor atoms around the N vacancy in the central plane have been found to be 0.04-0.07 Å (nearest U atoms from the neighboring layers are shifted in $z$-direction towards the vacancy), and do not exceed 0.025 Å for all the other atoms in the slab.

In contrast, the U vacancy results in much larger displacements of neighboring atoms, independently of its position. If the U vacancy is in the surface layer, then the atomic displacements of 0.3-0.32 Å are observed for the nearest N atoms. If the U vacancy lies in the central layer, then the nearest N atoms from this layer are displaced by 0.17 Å while the N atoms from the nearest layers are not shifted in $xy$ direction, being displaced by 0.15 Å towards the slab surface in the $z$-direction. Furthermore, the atomic displacements are weakly dependent on the slab thickness. The atomic displacements around the N and U vacancies in the UN bulk have been found to be -0.03 Å and 0.13 Å for N and U vacancies, respectively [15]. These values are close to those found in the present calculations for the vacancies in the central slab layer, which mimics the crystal bulk.

The finite slab-size effects caused by relatively large concentration of defects could be illustrated using the difference electron density redistribution $\Delta\rho(\mathbf{r})$. In Fig. 3, these redistributions are shown for N vacancies positioned at both the outer surface and central (mirror) planes of 5- and 7-layer slabs. Presence of two symmetrically positioned vacancies in the 5-layer slab induces their weak interaction across the slab (Fig. 3a) illustrated by appearance of an additional electron density around the N atoms in the central plane of the slab. Similarly, the vacancy in the mirror plane disturbs the atoms in the surface plane if thin slab contains only 5 layers (Fig. 3c). By increasing the slab thickness, we can avoid the effect of finite-slab size (Figs. 3b,d) which explains the stabilization of formation energies calculated for the 7-layer and thicker UN(001) slabs (Table 4).

The densities of states (DOS) are presented in Fig. 4 for perfect and defective 7-layer UN slab. The DOS for other slabs calculated in this study did not demonstrate additional effects and, thus, are not shown here. In accordance with previous bulk calculations [10, 15], the U(5$f$) electrons occupy the Fermi level (Fig. 4a). These electrons are relatively localized but still strongly hybridized with the N(2$p$) electrons. It confirms the existence of covalent bonding observed in the analysis of Bader charges (Table 2). The N(2$p$) states form a band of the width ~4 eV similar to that obtained in the bulk [10,15]. In contrast, the contribution of U(6$d$) electrons remains insensitive to the presence of vacancies as these are almost homogeneously distributed over a wide energy range including the conduction band.

The analysis of the average magnetic moment of U atoms ($m_{av}^{U}$) in the defective UN slabs is done too (Fig. 5). It decreases for both types of vacancies as a function of a number of layers in the slab, except for the U vacancy in the surface layer which remains almost unchanged. On the other hand, $m_{av}^{U}$ increases significantly when the U vacancy is located in the sub-surface and surface layers. In contrast to the U vacancies, $m_{av}^U$ for the slabs with the N vacancies are less sensitive to the position of defect. Moreover, the values of $m_{av}^U$ for the slabs with the N vacancies in the surface and sub-surface planes are practically identical.

**4. Conclusions**

In the present study, the basic properties of vacancies on the UN (001) surface were calculated from the first principles. In particular, the formation energies for U and N vacancies were determined using the two reference states, which included the energies of isolated atoms as well as atoms in the metallic α-U phase and $N_2$ molecule, respectively. The formation energies indicated a clear trend for segregation towards the surface (and probably, grain boundaries) as these energies for surface layer are noticeably smaller than those for sub-surface and central layers (although both latter are very close). However, the magnetic moments in the sub-surface and central layers differ significantly. We demonstrated also a considerable deviation of effective atomic charges from formal charges (caused by a covalent contribution to the U-N chemical bond). The obtained results will be used in the oncoming study of oxygen interaction with real (defective) UN surfaces, in order to understand the atomistic mechanism of UN oxidation.

**Acknowledgements**

This study was partly supported by the EC FP7 *F-BRIDGE* project and ESF project No.2009/0216/1DP/1.1.1.2.0/09/APIA/VIAA/044. D.B. gratefully acknowledges the doctoral studies support from the European Social Fund (ESF). The authors kindly thank R.A. Evarestov, P. Van Uffelen, and V. Kashcheyevs for fruitful discussions.

**Literature**

[1] P.D. Wilson (Ed.), The Nuclear Fuel Cycle, University Press, Oxford, 1996; www.gen-4.org .
[2] Y. Arai, M. Morihira, and T. Ohmichi, J. Nucl. Mater. 202 (1993) 70.
[3] R. Atta-Fynn and A. K. Ray, Phys. Rev. B 76 (2007) 115101.
[4] M.S.S. Brooks, J. Phys. F Metal Phys. 14 (1984) 639.
[5] M. Samsel-Czekała, E. Talik, P. de V. Du Plessis, R. Troć, H. Misiorek, and C. Sułkowski, Phys. Rev. B 76 (2007) 144426.
[6] M.S.S. Brooks and D. Glötzel, Physica 102B (1980) 51.
[7] E.A. Kotomin, Yu.A. Mastrikov, Yu. F. Zhukovskii, P. Van Uffelen, and V.V. Rondinella, Phys. stat. sol. (c) 4 (2007) 1193.
[8] R.A. Evarestov, M.V. Losev, A.I. Panin, N.S. Mosyagin, and A.V. Titov, Phys. stat. sol. (b) 245 (2008) 114.
[9] E.A. Kotomin and Yu.A. Mastrikov, J. Nucl. Mat. 377 (2008) 492.
[10] R.A. Evarestov, A.V. Bandura, M.V. Losev, E.A. Kotomin, Yu.F. Zhukovskii, and D. Bocharov, J. Comput. Chem. 29 (2008) 2079.
[11] R. Dovesi, V.R. Saunders, C. Roetti, R. Orlando, C.M. Zicovich-Wilson, F. Pascale, B. Civalleri, K. Doll, N.M. Harrison, I.J. Bush, Ph. D'Arco, and M. Llunell, CRYSTAL2006 User's Manual, Universita di Torino, Torino, 2006, http://www.crystal.unito.it/.

[12]  G. Kresse and J. Furthmüller, *VASP* the Guide, University of Vienna, 2009. http://cms.mpi.univie.ac.at/vasp/.
[13] Yu.F. Zhukovskii, D. Bocharov, E.A. Kotomin, R.A. Evarestov, and A.V. Bandura, Surf. Sci. 603 (2009) 50.
[14] Yu.F. Zhukovskii, D. Bocharov, and E.A. Kotomin, J. Nucl. Mater. 393 (2009) 504.
[15] E.A. Kotomin, R.W. Grimes, Yu.A. Mastrikov, and N.J. Ashley, J. Phys.: Condens. Matter 19 (2007) 106208.
[16] E.A. Kotomin, Yu.A. Mastrikov, S. Rashkeev, and P. van Uffelen, J. Nucl. Mat. 393 (2009) 292.
[17] G. Kresse and J. Furthmüller, Comput. Mater. Sci. 6 (1996) 15.
[18] G. Kresse and D. Joubert, Phys. Rev. A 59 (1999) 1758.
[19] J.P. Perdew, J.A. Chevary, S.H. Vosko, K.A. Jackson, M.R. Pederson, D.J. Singh, and C. Fiolhais, Phys. Rev. B 46 (1992) 6671.
[20] H.J. Monkhorst and J.D. Pack, Phys. Rev. B 13 (1976) 5188.
[21] M. Methfessel and A.T. Paxton, Phys. Rev. B 40 (1989) 3616.
[22] H-J. Matzke, Science of Advanced LMFBR Fuels, North Holland, Amsterdam, 1986.
[23] R. Bader, Atoms in Molecules: A Quantum Theory, Oxford University Press, New York, 1990.
[24] G. Henkelman, A. Arnaldsson, and H. Jónsson, Comput. Mater. Sci. 36 (2006) 354.
[25] J. Akella, S. Weir, J. M. Wills, P. Söderlind, J. Phys.: Condens. Matter. 9 (1997) L549.
[26] C. G. Van de Walle, J. Neugebauer, J. Appl. Phys. 95 (2004) 3851.
[27] P. Söderlind, Phys. Rev. B 66 (2002) 085113.
[28] B. Dorado, M. Freyss, and G. Martin, Eur. Phys. J. B 69 (2009) 203.
[29] M. Freyss, Phys. Rev. B 81 (2010) 014101.
[30] D.R. Lide (ed.), CRC Handbook of Chemistry and Physics, 88th Edition, CRC Press (2007-2008).
[31] M. Iwasawa, Y. Geng, Y. Kaneta, T. Ohnuma, H.-Y. Geng, and M. Kinoshita, Mater. Trans. 47 (2006) 014101.
[32] NIST Chemistry Web-book (2010); http://www.webbook.nist.gov/chemistry/.

Table 1. Surface energies $E_{surf}$ (J·m$^{-1}$) for calculations with relaxed and unrelaxed atomic spins as well as averaged magnetic moment (in μ$_B$) of U atom for the defect-free UN (001) surface.

| Number of atomic planes | $E_{surf}$ spin unrelaxed slab | $E_{surf}$ spin-relaxed slab | $\mu_{av}$ |
|---|---|---|---|
| 5 | 1.69 | 1.44 | 1.57 |
| 7 | 1.70 | 1.37 | 1.44 |
| 9 | 1.70 | 1.29 | 1.37 |
| 11 | 1.69 | 1.22 | 1.33 |

Table 2. Atomic Bader charges on a defect-free surface.

| Atom | Number of atomic layers | | | |
|---|---|---|---|---|
| | 5 | 7 | 9 | 11 |
| Surface U | 1.68 | 1.74 | 1.68 | 1.72 |
| Sub-surface U | 1.67 | 1.63 | 1.63 | 1.67 |
| U in central (mirror) plane | 1.69 | 1.72 | 1.65 | 1.66 |
| Surface N | -1.65 | -1.67 | -1.67 | -1.68 |
| Sub-surface N | -1.68 | -1.70 | -1.70 | -1.67 |
| N in central (mirror) plane | -1.74 | -1.65 | -1.65 | -1.63 |

Table 3. Atomic displacements $\Delta z$(Å)$^*$ for defect-free surface (spin-relaxed calculations).

| Number of atomic planes | U atom displacements | | N atom displacements | |
|---|---|---|---|---|
| | Surface | Sub-surface | Surface | Sub-surface |
| 5 | -0.050 | -0.012 | 0.023 | 0.023 |
| 7 | -0.046 | -0.009 | 0.024 | 0.028 |
| 9 | -0.047 | -0.011 | 0.024 | 0.028 |
| 11 | -0.047 | -0.011 | 0.025 | 0.031 |

$^*$ negative sign means an inward atomic displacement towards the slab center

Table 4. The vacancy formation energies (in eV) for the two reference states (see the text for details).

| Layer | Number of atomic planes in slab and supercell extension (in brackets) | Reference I, Eqs. (1a)–(2)[a] | | Reference II, Eqs. (1a), (1b), (3a) and (3b)[b] | |
|---|---|---|---|---|---|
| | | U | N | U | N |
| Surface layer | 5 (2×2) | 8.63 | 8.84 | 1.46 | 3.70 |
| | 7(2×2) | 8.61 | 8.84 | 1.44 | 3.70 |
| | 9(2×2) | 8.61 | 8.84 | 1.44 | 3.71 |
| | 11(2×2) | 8.60 | 8.85 | 1.43 | 3.71 |
| | 5(3×3) | 8.51 | 8.78 | 1.34 | 3.64 |
| | 7(3×3) | 8.47 | 8.78 | 1.30 | 3.65 |
| Sub-surface layer | 5(2×2) | 10.31 | 9.38 | 3.14 | 4.25 |
| | 7(2×2) | 10.29 | 9.46 | 3.12 | 4.33 |
| | 9(2×2) | 10.26 | 9.46 | 3.09 | 4.33 |
| | 11(2×2) | 10.26 | 9.46 | 3.09 | 4.33 |
| | 7(3×3) | 10.18 | 9.47 | 3.01 | 4.34 |
| Central (mirror) layer[c] | 5(2×2) | 10.20 | 9.48 | 3.03 | 4.34 |
| | 7(2×2) | 10.36 | 9.57 | 3.19 | 4.43 |
| | 9(2×2) | 10.34 | 9.55 | 3.17 | 4.42 |
| | 11(2×2) | 10.39 | 9.56 | 3.22 | 4.42 |
| | 7(3×3) | 10.23 | 9.55 | 3.06 | 4.42 |

[a] reference energies I equal to -4.10 eV for U atom and -3.17 eV for N atom,
[b] reference energies II equal to -11.28 eV for U atom and -8.30 eV for N atom,
[c] defect formation energies for UN bulk using reference I are 9.1-9.7 eV for N vacancy and 9.4-10.3 for U vacancy [15].

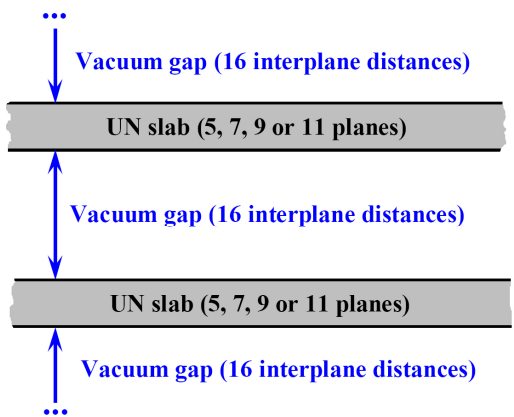 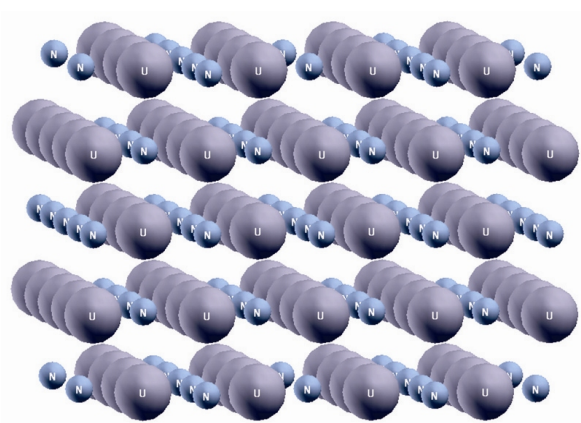

*Fig. 1.* Cross-section of UN (001) slabs.

*Fig. 2 (Color online).* 5-layer slab containing the two-sided surface N vacancies with a 2×2 periodicity.

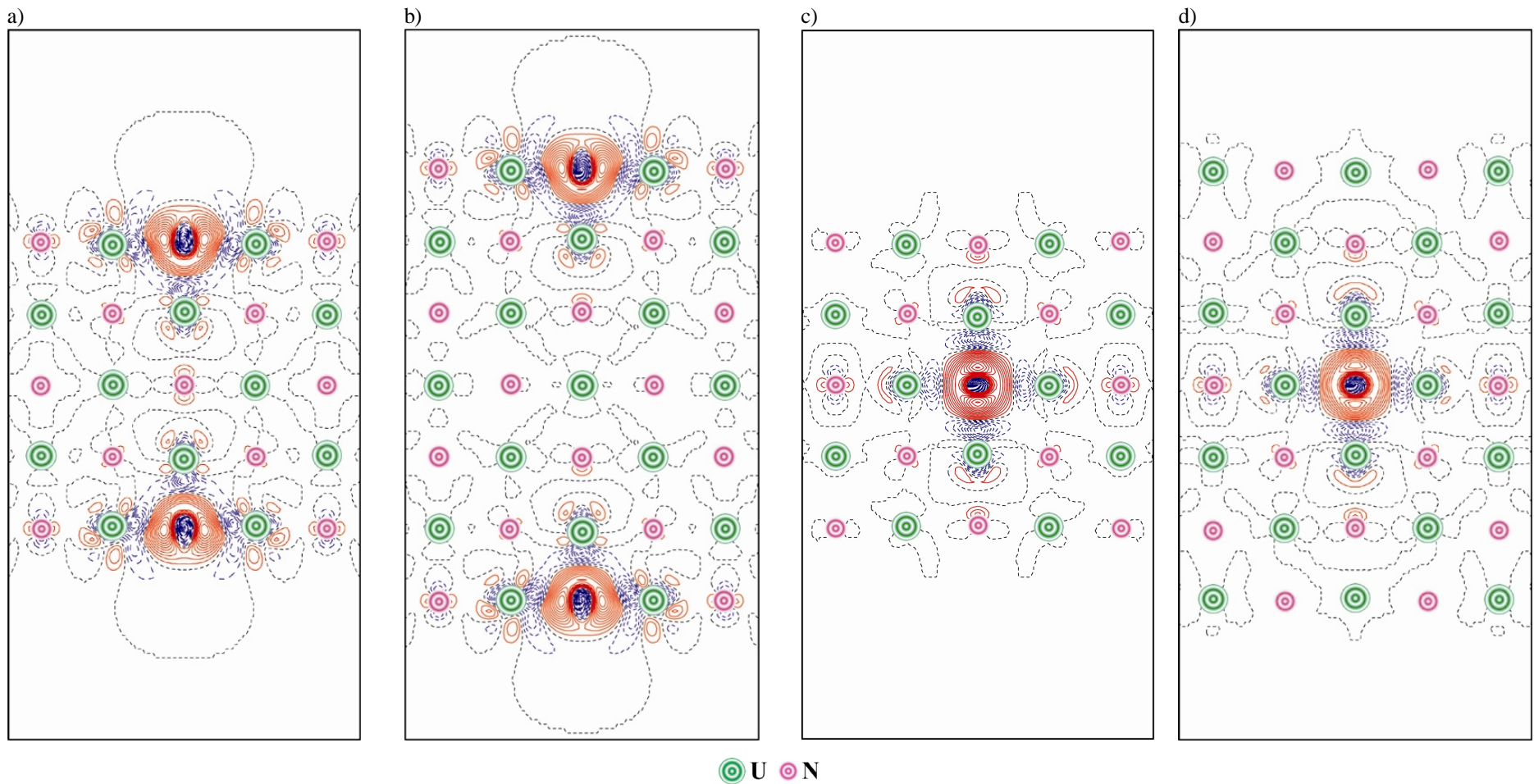

*Fig. 3 (Color online).* 2D sections of the electron density redistributions around the nitrogen vacancies in five- and seven-layer UN(001) slabs with 2×2 supercell extension defined as the total electron density of defected surface minus a superposition of the electron densities for both perfect surface and isolated atom in the regular position on the surface: a) N vacancy in a surface plane, five-layer slab, b) the same, 7-layer slab, c) N vacancy in a central plane, five-layer slab, b) the same, 7-layer slab. Solid (red) and dashed (blue) isolines correspond to positive and negative electron density, respectively. (For interpretation of the references to color in this Figure legend, the reader is referred to the web version of this article). Isodensity increment is 0.25 e a.u.$^{-3}$.

a)

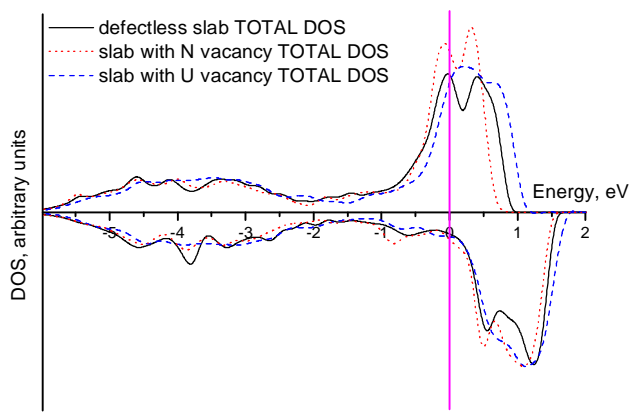

b)

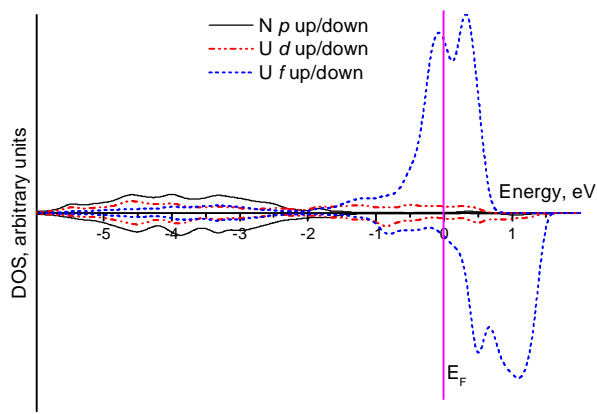

c)

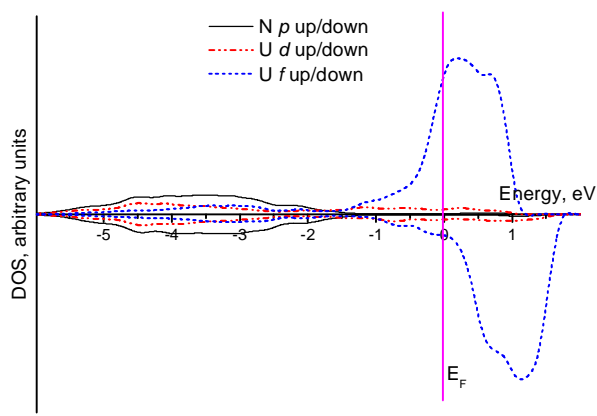

*Fig. 4 (Color online).* The total and projected DOSs of 7-layer UN(001) slab (2×2 supercell for vacancy-containing models): a) total DOS of defective and defect-free surfaces, b) projected DOSs for the surface containing N vacancies, c) projected DOSs for the surface containing U vacancies.

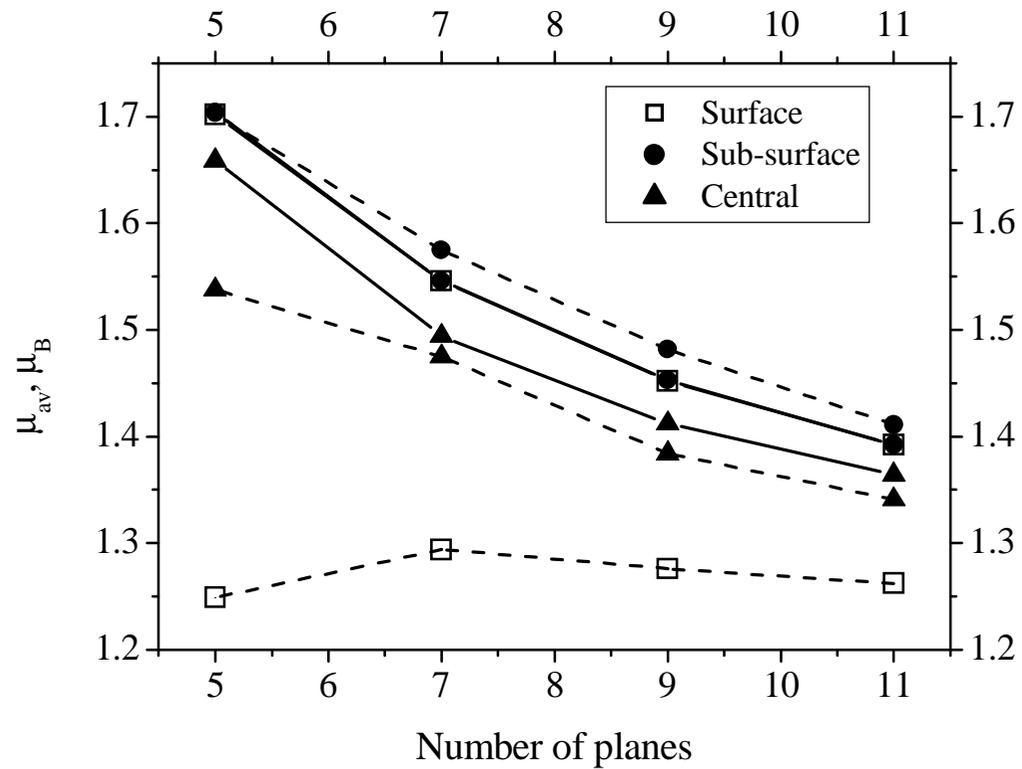

*Fig. 5.* The average U magnetic moment $\mu_{av}$ (in $\mu_B$) in the slab as a function of a number of planes. The dashed curves correspond to U vacancy whereas the solid curves describe the N vacancy.